\documentclass{aastex}
\usepackage{spr-astr-addons}
\usepackage{url}
\usepackage{color}

\RequirePackage{color}

\begin{document}

\title{Ohm's Law for Plasma in General Relativity and Cowling's Theorem}
\slugcomment{}
\shorttitle{Ohm Law for Plasma in GR} \shortauthors{Bahodir
Ahmedov}

\author{Bahodir B. Ahmedov\altaffilmark{1}}

\altaffiltext{1}{Applied Mathematics and Informatics Department,
Uzbekistan National University, Tashkent 100174, Uzbekistan}

\begin{abstract}
The general-relativistic Ohm's law for a two-component plasma
which includes the gravitomagnetic force terms even in the case of
quasi-neutrality has been derived. The equations that describe the
electromagnetic processes in a plasma surrounding a neutron star
are obtained by using the general relativistic form of Maxwell
equations in a geometry of slow rotating gravitational object. In
addition to the general-relativistic effect first discussed by
Khanna $\&$ Camenzind (1996) we predict a mechanism of the
generation of azimuthal current under the general relativistic
effect of dragging of inertial frames on radial current in a
plasma around neutron star. The azimuthal current being
proportional to the angular velocity $\omega$ of the dragging of
inertial frames can give valuable contribution on the evolution of
the stellar magnetic field if $\omega$ exceeds $2.7\times 10^{17}
(n/\sigma) \textrm{s}^{-1}$ ($n$ is the number density of the
charged particles, $\sigma$ is the conductivity of plasma). Thus
in general relativity a rotating neutron star, embedded in plasma,
can in principle generate axial-symmetric magnetic fields even in
axisymmetry. However, classical Cowling's antidynamo theorem,
according to which a stationary axial-symmetric magnetic field can
not be sustained against ohmic diffusion, has to be hold in the
general-relativistic case for the typical plasma being responsible
for the rotating neutron star.
\end{abstract}

\keywords{MHD; plasmas; general relativity; magnetic field.}

\section{Introduction}

The discovery of pulsars, characterized by the emission of radio
pulses at regular intervals has lead to the belief that pulsars
are compact rotating neutron stars with extremely large frozen-in
magnetic fields. The high conductivity of the stellar interior
ensures conservation of magnetic flux during collapse. For this
reason, it is believed the strength of initial magnetic field will
increase quadratically with the decrease of the linear dimensions
and reach, as a rule, up to $10^{12} \textrm{G}$ in the final
state for the young neutron star. The time evolution of such
strong magnetic field has been the subject of much discussion over
the years and the reviews (e.g.~\cite{r09}, \cite{g09},
\cite{b02}, Lamb 1991; Chanmugam 1992; Phinney $\&$ Kulkarni 1994)
provide the present understanding on the decay of magnetic fields
in isolated neutron stars.

Recent observation of pulsars and their
statistical analysis seem to imply that the evolution of magnetic
fields of isolated pulsars is still relatively open and the evidence
supporting that they do not undergo  or undergo significant magnetic field
decay is not conclusive due to large error bars. However it is well-known,
in general, a dynamo, when the motions of a conducting fluid are able to
sustain or increase a stellar magnetic field, needs.
But for a number of simple geometries no dynamo is possible and
according to the famous Cowling's antidynamo theorem (Cowling 1934) the
stationary axial-symmetric magnetic field can not be self maintained.

The situation may be different in a general-relativistic context.
That is why the kinematic evolution of axisymmetric magnetic and
electric fields has been recently investigated in Kerr geometry
(Khanna $\&$ Camenzind; Brandenburg 1996; Nunez 1996, 1997).
Interesting other types of fast dynamo mechanisms, based on
stretching flux tubes in Riemannian conformal manifolds, have been
also recently obtained by Garcia de Andrade (2007, 2008). However
it was found in (Khanna $\&$ Camenzind; Brandenburg 1996,
\cite{mz06}) that a magnetic field can not be also sustained
against ohmic diffusion in general relativistic case as in flat
space-time one. No support was found for the possibility that the
general relativistic effects could lead to self-excited
axisymmetric solutions. Nunez (1997) states that the
gravitomagnetic potential in the Kerr metric couples the equations
of the magnetic flux and current, rendering invalid the argument
used in the proof of Cowling's antidynamo theorem. In this respect
the magnetic field evolution, especially general relativistic
contribution to it, around and in a supermassive stars is
extremely significant component in the recent models of pulsars
and rotating neutron stars.

Electrodynamics in a four-dimensional spacetime ''feels'' inertial
and gravitational effects via the metric dependent constitutive
(spacetime and material) relations. For conductors, this is
manifest in the covariant generalization of Ohm's law see e.g.
\cite{a99b}. Relativistic version of the generalized Ohm's law for
plasma can be found e.g. in Ardavan (1976), Blackman $\&$ Field
(1993), Gedalin (1996), Khanna (1998), Kremer \& Patsko (2003),
Meier (2004), \cite{kt08}, \cite{k09}.

In this paper the classical derivation of Cowling's theorem is
repeated using general relativistic electromagnetic equations
governing two-component plasma in the background geometry of
stationary gravitational body. This set of equations has general
relativistic Ohm's law for plasma including new
general-relativistic gravitomagnetic terms, which may be
fundamental for the study the generation and evolution of stellar
magnetic field. We obtain a nonvanishing general relativistic
expression for the circulation of electric current, which is due
to the general relativistic frame dragging effect on the radial
electric current flowing in plasma in the vicinity of a rotating
neutron star. Thus we show, that in axisymmetry, the
gravitomagnetic effects can drive currents and generate magnetic
fields, even without taking into account the possibility of
turbulence and $\alpha$ -- dynamo effect. But from our evaluations
for the typical astrophysical plasma the generated magnetic field
may be extremely weak in order to be taken into account.

\section{Ohm's law for plasma in general relativity}

One may consider the electrons and ions in plasma as separate
fluids which are interacting with each other through collisions.
This two-fluid model is also essential for deriving the general
relativistic Ohm's law for plasma and for describing the different
effects being responsible for the generation of the
electromagnetic fields. To find this law we consider the equations
describing the motion of the individual components of the plasma,
i.e. the linearized equations of motion of electrons and ions
\begin{eqnarray}
\label{1} c^2u^\alpha_{(e);\sigma}u^\sigma_{(e)}=
-\frac{e}{m}F^{\alpha\beta}u_{(e)
\beta}-\nu_1\left(v^\alpha_{(e)}-v^\alpha_{(i)}\right) \nonumber\\
-\frac{\Lambda^{-1/2}}{mN_e}\stackrel{\perp}{\nabla}_\alpha
\tilde{p}_e \ , \\
\label{2} c^2u^\alpha_{(i);\sigma}u^\sigma_{(i)}=
\frac{e}{M}F^{\alpha\beta}u_{(i)
\beta}-\nu_2\left(v^\alpha_{(i)}-v^\alpha_{(e)}\right) \nonumber\\
-\frac{ \Lambda^{-1/2}}{M_iN_i}\stackrel{\perp}{\nabla}_\alpha
\tilde{p}_i \ ,
\end{eqnarray}
where $\nu_1$ and $\nu_2$ are the collision frequencies, $m$ and
$M_i$ are the mass of electron and ion, the subscripts $i$ and $e$
denote ion and electron quantities, respectively, semicolon is the
covariant derivative, c is the speed of light,
$\left(\Lambda^{1/2}p_e\right) = \tilde{p}_e$,  the term
$\Lambda^{-1/2}\stackrel{\perp}{\nabla}_\alpha
\left(\Lambda^{1/2}p_{e,i}\right)$ is due to the change of the
pressure $p_{e,i}$ of electron and ion components and
$\stackrel{\perp}{\nabla}_\alpha$ denotes the spatial part of
covariant derivative, $F_{\alpha\beta}$ is the tensor of
electromagnetic field.

The gravitational field is assumed to be
stationary that is space-time metric $g_{\alpha\beta}$ admits a
timelike Killing vector $\xi^\alpha_{(t)}$
that is
${\it\L}_{\xi_t} g_{\alpha\beta}=0$ ({\it\L}$_{\xi_t}$ denotes the Lie
derivative with respect to $\xi^\alpha_{(t)}$,
$\Lambda =-\xi^\alpha_{(t)}\xi_{(t)\alpha}$). The gravitational field,
represented by the metric tensor, is assumed to be generated by outside
gravitational source. The plasma itself is expected to generate a much
weaker gravitational field.

After doing some algebraic transformations equations (\ref{1}) and
(\ref{2}) can be written in the form:
\begin{eqnarray}
\label{3}
c\partial_Tv^\alpha_{(e)}=-\frac{e}{m}F^{\alpha\beta}u_\beta-\frac{e}{mc}
\left(F^{\alpha\sigma}+F^{\rho\sigma}u_\rho
u^\alpha\right)v_{(e)\sigma}
\nonumber\\
-\nu_1\left(v_{(e)}^\alpha -v_{(i)}^\alpha \right)-c^2w^\alpha -
2cv^\beta_{(e)}A^\alpha_{.\beta}
-\frac{\Lambda^{-1/2}}{mN_e}\stackrel{\perp}{\nabla}_\alpha
\tilde{p}_e \ , \\
\label{4}
c\partial_Tv^\alpha_{(i)}=\frac{e}{M_i}F^{\alpha\beta}u_\beta+\frac{e}{M_ic}
\left(F^{\alpha\sigma}+F^{\rho\sigma}u_\rho u^\alpha\right)v_{(i)\sigma}
\nonumber\\
- \nu_2\left(v_{(i)}^\alpha -v_{(e)}^\alpha \right)-c^2w^\alpha -
2cv^\beta_{(i)}A^\alpha_{.\beta}-
\frac{\Lambda^{-1/2}}{M_iN_i}\stackrel{\perp}{\nabla}_\alpha
\tilde{p}_i \ .
\end{eqnarray}
Here $u_\beta$ is the four-velocity of the proper frame,
$u_{\mu;\nu}=A_{\mu\nu}-D_{\mu\nu}+w_\mu u_\nu$,
$A_{\beta\alpha}=u_{[\alpha ,\beta]}+u_{[\beta}w_{\alpha]}$ is the
relativistic rate of rotation, $w_\alpha=u_{\alpha;\beta}u^\beta$
is the absolute acceleration, $[\cdots]$ denotes the
antisymmetrization, $\partial_T$ denotes the time derivative
(Vladimirov 1982; Antonov et al 1978),
$D_{\mu\nu}=\partial_Th_{\mu\nu}/2$\footnote{It is assumed that
the spacetime is stationary where $D_{\mu\nu}=0$.} is the tensor
of deformation velocities, $h_{\mu\nu}=g_{\mu\nu}+u_\mu u_\nu$;
the relative velocities of electrons and ions are
\begin{eqnarray}
\label{5}
v^\alpha_{(e)}/c=\sqrt{1-v_{(e)}^2/c^2}u^\alpha -u^\alpha_{(e)} \ , \\
\label{6}
v^\alpha_{(i)}/c=\sqrt{1-v_{(i)}^2/c^2}u^\alpha -u^\alpha_{(i)} \ .
\end{eqnarray}

The mean velocity $v^\alpha$, the current density $\jmath^\alpha$
and the mass density $\rho_m$ are
\begin{eqnarray}
\label{7}
v^\alpha=\frac{\left(M_iN_iv^\alpha_{(i)}+mN_ev^\alpha_{(e)}\right)}
{M_iN_i+mN_e} \ , \\
\label{8}
\jmath^\alpha=e\left(N_iv^\alpha_{(i)}-N_ev^\alpha_{(e)}\right) \ , \\
\rho_m=M_iN_i+mN_e \ .
\end{eqnarray}

For simplicity we here assume that $N_e=N_i=N$ (i.e. the charge neutrality)
and since $M_i\gg m$, we find
that
\begin{eqnarray}
\label{9}
v^\alpha=v^\alpha_{(i)}+\frac{m}{M_i}v^\alpha_{(e)} \ , \nonumber\\
\label{10}
\jmath^\alpha=Ne\left\{v^\alpha-\left(\frac{m}{M_i}+1\right)v^\alpha_{(e)}\right\}
\ .
\end{eqnarray}
Hence
\begin{eqnarray}
\label{11}
v^\alpha_{(e)}=-\frac{j^\alpha}{Ne}+v^\alpha \ ,\\
\label{12} v^\alpha_{(i)}=v^\alpha+\frac{m}{\rho_m e}\jmath^\alpha
\ .
\end{eqnarray}

From putting the derived expressions for $v^\alpha _{(i)}$ and $v^\alpha_
{(e)}$ (\ref{11}) and (\ref{12}) into formulae (\ref{3}) and (\ref{4}),
we obtain
\begin{eqnarray}
\label{13}
&&\partial_T\left(v^\alpha-\frac{\jmath^\alpha}{Ne}\right)=
-\frac{e}{mc}F^{\alpha\beta} u_\beta-\frac{\nu_1}{c}
\left(v^\alpha_{(e)}-v^\alpha_{(i)}\right)\nonumber\\
&&-\frac{e}{mc^2}\left(F^{\alpha\sigma}+ F^{\rho\sigma}u_\rho
u^\alpha\right)
\left(v_\sigma-\frac{\jmath_\sigma}{Ne}\right)-cw^\alpha
\nonumber\\
&&-
2\left(v^\beta-\frac{\jmath^\beta}{Ne}\right)A^\alpha_{.\beta}-
\frac{\Lambda^{-1/2}}{cmN_e}\stackrel{\perp}{\nabla}_\alpha
\tilde{p}_e \ , \\
\label{14} &&\partial_T\left(v^\alpha+\frac{m\jmath^\alpha}{\rho_m
e}\right)=\frac{e}{M_ic}F^{\alpha\beta} u_\beta- \frac{\nu_2}{c}
\left(v^\alpha_{(i)}-v^\alpha_{(e)}\right)\nonumber\\
&&+\frac{e}{M_ic^2}\left(F^{\alpha\sigma}+ F^{\rho\sigma}u_\rho
u^\alpha\right) \left(v_\sigma +\frac{m\jmath_\sigma}{\rho_m
e}\right)-cw^\alpha
\nonumber\\&&-2\left(v^\beta+\frac{m\jmath^\beta}{\rho_m
e}\right)A^\alpha_{.\beta}
-\frac{\Lambda^{-1/2}}{cM_iN_i}\stackrel{\perp}{\nabla}_\alpha
\tilde{p}_i \ .
\end{eqnarray}

In order to obtain the generalized Ohm's law for the plasma as a single
fluid we combine the equations of motion and substract (\ref{13}) from
(\ref{14}), and get approximately
\begin{eqnarray}
\label{15} F^{\alpha\beta}u_{\beta}+\frac{1}{c}
\left(F^{\alpha\sigma} +F^{\rho\sigma}u_\rho u^\alpha
\right)v_\sigma= -\frac{\Lambda^{-1/2}}{Ne}
\stackrel{\perp}{\nabla}_\alpha\tilde{p}_e\nonumber\\
+\frac{\jmath^\alpha}{\sigma} -
\frac{1}{Nec}\left(F^{\alpha\sigma}+ F^{\rho\sigma}u_\rho u^\alpha
\right)\jmath_\sigma \nonumber\\-\frac{mc}{Ne^2}
\partial_T \jmath^\alpha+\frac{2mc}{Ne^2}\jmath^\beta A^\alpha_{.\beta}.
\label{eq:ohmgen}
\end{eqnarray}
Here $\nu=\nu_1+\nu_2$, $\sigma=\frac{Ne^2}{m\nu}$ is the
electrical conductivity in the presence of a constant electric
field and zero magnetic field.

We follow the description of the generalized Ohm's law (\ref{15})
from the review paper~\cite{bs05}. The first term on the right
hand side  of equation (\ref{15}), being produced by the electron
pressure gradient, is the Biermann battery term~\cite{b50}. For
example it may provide the source term for the thermally generated
electromagnetic fields~\cite{mr62}. The next two terms on the
right hand side  of equation (\ref{15}) are the usual Ohmic term
and the Hall electric field, which arises due to a nonvanishing
Lorentz force.  The next term on the right hand side  is the
inertial term, which can be neglected if the macroscopic time
scales are large compared to the plasma oscillation periods. And
finally the last term on the right hand side of equation
(\ref{15}) appears due to the Coriolis force and dragging of
inertial frames effects on the conduction current.

In neutron stars, the presence of strong magnetic fields, could
make the Hall term important, especially in their outer regions,
where there are also strong density gradients. The Hall effect in
neutron stars can lead to magnetic fields undergoing a turbulent
cascade~\cite{gr92}. By analogy with the vorticity equation in
ordinary hydrodynamics,~\cite{gr92}
 conjectured that the transfer of
magnetic energy from large to small scales proceeds in a similar
way to ordinary turbulence. However, the analogy of the Hall
induction equation with the vorticity equation is not complete,
and the conjecture remained to be confirmed by multidimensional
numerical simulations. It can also lead to a nonlinear steepening
of field gradients~\cite{vco00} for purely toroidal fields, and
hence to enhanced magnetic field dissipation. \cite{vco00}
proposed a mechanism for the fast dissipation of magnetic field
based on the Hall drift in stratified media. They correctly
pointed out that Hall currents are able to create current sheets
(which are sites for efficient dissipation) and that the evolution
of the toroidal field resembles the Burgers equation. The same
Burgers-like equation is applicable even to non-stratified media,
but in a spherical shell (\cite{pg07},\cite{pg10}), in which the
Hall term in the induction equation tends to create current sheets
instead of ordinary turbulence. \cite{rg02} showed by a linear
analysis that, in a one-component (electron) plasma, a large-scale
background magnetic field may become unstable to smaller scale
perturbations. This Hall-drift induced instability occurs when the
magnetization parameter is high and the background field has
enough curvature. Since these conditions may be realized in the
crust of a neutron star, the problem of the Hi became interesting
not only from the magnetohydrodynamic point of view but also for
the astrophysics community. Although the Hall-drift is a
non-dissipative process, the growth of small-scale magnetic field
components modifies the overall magnetic field structure and opens
the possibility of more rapid field decay than pure ohmic
dissipation would predict.

For a stationary plasma $v_\sigma=0$ and in the steady state
$\partial_T j^\alpha=0$, this equation becomes
\begin{eqnarray}
\jmath^\alpha=\frac{Ne^2}{m\nu}\{F^{\alpha\beta}u_{\beta}
+\frac{\Lambda^{-1/2}}{Ne}\stackrel{\perp}{\nabla}_\alpha
\tilde{p}_e \}-\nonumber\\
-\frac{e}{m\nu c}\left(F^ {\alpha\sigma}+F^{\rho\sigma}u_\rho
u^\alpha \right)\jmath_\sigma + \frac{2c}{\nu}\jmath^\beta
A^\alpha_{.\beta}.
\end{eqnarray}
It can be written as
\begin{eqnarray}
\frac{\jmath_\alpha}{\sigma}=F_{\alpha\beta}u^{\beta}-R_H \left(F_
{\alpha\sigma}+F_{\rho\sigma}u^\rho u_\alpha \right)\jmath^\sigma
+
\nonumber\\
R_{gg}\jmath^\beta A_{\alpha\beta}+\frac{\Lambda^{-1/2}}{Ne}
\stackrel{\perp}{\nabla}_\alpha\tilde{p}_e. \label{eq:ohm's}
\end{eqnarray}
This is the generalized Ohm's law for two-component plasma in general
relativity. Here
\begin{equation}
R_H=\frac{1}{Nec}, \quad R_{gg}=\frac{2mc}{Ne^2},
\end{equation}
obviously  $R_H$ is the Hall constant, $R_{gg}$ is the parameter
for the plasma called as galvano-gravitomagnetic one.

Khanna (1998) has derived the general relativistic Ohm's law for
two-component plasma and concluded that it has no new terms as
compared with special relativity in the limit of quasi-neutral
plasma. From our point of view the gravitomagnetic terms did not
appear in Ohm's law most probably as a consequence of the
magnetohydrodynamic approximation used in (Khanna 1998). The first
two terms in the right hand side of equation~(\ref{eq:ohm's}) are
standard classical terms which include the general relativistic
contributions. The third term in the right hand side of
equation~(\ref{eq:ohm's}) has been discussed for the conduction
current in conductors (Ahmedov 1998, 1999a,b). It has purely
relativistic nature and is caused by the effect of gravitomagnetic
force on the electric current flowing in the plasma. Also it has
recently been obtained for plasma by~\cite{kt08}, see to the third
term under the square bracket in the right hand side of their
equation (56).

\section{Space charge density in two-component plasma}

As a consequence of (\ref{eq:ohm's}) for a plasma without
conduction current $\jmath^\alpha=0$, electric field
$E^\alpha=F^{\alpha\beta}u_\beta=
-\stackrel{\perp}{\nabla}_\alpha\tilde{p}_e$. This electric field
inside a plasma in a gravitational field is a well-known
phenomenon in the magnetosphere and in stellar interiors. In the
plasma magnetosphere, gravity acts strongly on ion constituent
while the electron constituent tends to escape to infinity. An
upward-directed electric field is set up that prevents the
electrons from escaping, thus charge neutrality is maintained
within magnetospheric medium. In the magnetosphere, the strength
of this upward-directed electric field depends on the mass of ion
constituent and on the electron temperature.

The number density of each particle species varies exponentially as
$-\tilde\mu /kT$ (Ehlers 1971), where $\tilde\mu=\Lambda^{1/2}\mu$ is
the gravito-electro-chemical potential including the rest mass energy, $k$
is Boltzmann's constant, and $T$ is the temperature.

To
preserve electrical quasi-neutrality the variation of the densities of
ions and electrons with height must be essentially the same for each
constituent, giving therefore
\begin{equation}
\tilde\mu_i/kT_i=\tilde\mu_e/kT_e
\end{equation}
and consequently
\begin{equation}
M_ic^2w_\alpha+eE_\alpha/T_i=mc^2w_\alpha-eE_\alpha/T_e,
\end{equation}
where $E_\alpha$ is the electric field required to maintain charge
neutrality
in the presence of gravitational field. The field $E_\alpha$ is then, for
single charged ions,
\begin{equation}
E_\alpha=\frac{w_\alpha\left(mc^2T_i-M_ic^2
T_e\right)}{e\left(T_e+T_i\right)},
\end{equation}
or for physically realisable case where $T_e$ is comparable with $T_i$ and
$M_i\gg m$
\begin{equation}
E_\alpha=\frac{M_ic^2w_\alpha}{e}\frac{T_e}{T_i+T_e} \ .
\label{eq:ief}
\end{equation}

We see that the induced electric field in a plasma is, to an excellent
approximation, principally a function of the ion mass and electron
temperature. In the case of an isothermal plasma composed entirely of
electrons and positrons, electric field $E_\alpha=0$.

If we suppose that the material relations between inductions and fields
have linear character i.e.
\begin{eqnarray}
H_{\alpha\beta}=\frac{1}{\mu}F_{\alpha\beta}+\frac{1-\epsilon\mu}{\mu}
\left(u_\alpha F_{\sigma\beta} -u_\beta F_{\sigma\alpha}\right)u^\sigma \ ,
\\
F_{\alpha\beta}=\mu H_{\alpha\beta}+\frac{\epsilon\mu -1}{\epsilon}
\left(u_\alpha H_{\sigma\beta} -u_\beta H_{\sigma\alpha}\right)u^\sigma
\end{eqnarray}
 and use the generalized Ohm's law for plasma
then one can easily derive the general formula for the space
charge distribution inside plasma
\begin{eqnarray}
\rho_0&=&\frac{\epsilon\mu
R_H}{c}\jmath^2+\frac{1}{4\pi}\Bigg\{\left(\frac{\epsilon}
{\sigma}\jmath^\alpha\right)_{;\alpha}- \epsilon
w^\alpha\frac{\Lambda^{-1/2}}{Ne}
\stackrel{\perp}{\nabla}_\alpha\tilde{p}_e
\nonumber\\
&&+\left[\epsilon^2\mu R_H\left(\frac{1}{\sigma}\jmath^2+
\frac{\Lambda^{-1/2}}{Ne}\jmath^\nu\stackrel{\perp}{\nabla}_\nu
\tilde{p}_e\right)u^\alpha \right]_{;\alpha}
\nonumber\\
&&-\epsilon R_{gg}A_{\alpha\beta} w^\alpha\jmath^\beta+
g^{\alpha\beta}\left(\epsilon R_{gg}\jmath^\nu
A_{\alpha\nu}\right)_{;\beta}-
\frac{\epsilon}{\sigma}w^\alpha\jmath_\alpha \nonumber\\
&&+ g^{\alpha\beta}\left(\epsilon \frac{\Lambda^{-1/2}}{Ne}
\stackrel{\perp}{\nabla}_\alpha\tilde{p}_e\right)_{;\beta}
\nonumber\\
&&+H^{\alpha\beta}[A_{\beta\alpha}+\epsilon\mu
R_Hw_\alpha\jmath_\beta +\left(\epsilon\mu
R_H\jmath_\alpha\right)_{,\beta}]\Bigg\}
\end{eqnarray}
from the general relativistic Maxwell equations
\begin{eqnarray}
e^{\alpha\beta\mu\nu} F_{\beta\mu ,\nu}= 0,\quad
{H^{\alpha\beta}}_ {;\beta}= \frac{4\pi}{c}J^\alpha,\quad
J^\alpha= c\rho_0 u^\alpha +\jmath^\alpha. \label{eq:max}
\end{eqnarray}
Here $H_{\alpha\beta}$ is the tensor of electromagnetic induction,
$\epsilon$ and $\mu$ are the parameters for the plasma.

Even in the case when there are no any currents flowing in plasma
($\jmath^\alpha=0$) the nonvanishing space charge
\begin{eqnarray}
\rho_0&=&\frac{1}{4\pi}\{g^{\alpha\beta}\left(\epsilon\frac{\Lambda^{-1/2}}{Ne}
\stackrel{\perp}{\nabla}_\alpha\tilde{p}_e\right)_{;\beta}
\nonumber\\ &&- \epsilon w^\alpha\frac{\Lambda^{-1/2}}{Ne}
\stackrel{\perp}{\nabla}_\alpha\tilde{p}_e+
H^{\alpha\beta}A_{\beta\alpha}\}
\end{eqnarray}
appears and is a sum of two contributions: one is due to the inner
electric field (\ref{eq:ief}) discussed above and second one is
the general-relativistic generalization of the Goldreich-Julian
charge density (Goldreich \& Julian 1969) in plasma magnetosphere
of rotating magnetized neutron star.

\section{On Cowling's theorem in stationary gravitational field}

According to the Cowling's theorem (Cowling 1934), steady plasma
motions can not maintain a magnetic field that is confined to a
finite region of space and possesses axial symmetry.  Our task is
to reformulate this theorem for the case when external
gravitational field exists, i.e. to generalize to the general
relativistic context the simple version of this theorem presented
for example in~\cite{c98}.

The metric of an asymptotically flat, stationary, axially
symmetric spacetime around a rotating gravitating body (see, e.g.
Landau \& Lifshitz (1975)) is considered. In spherical polar
coordinates $x^0 = ct, x^1 = r, x^2 = \theta$ and $x^3=\varphi$,
we have
\begin{eqnarray}
ds^2 =-e^{2\Phi(r)} \left(cdt\right) ^2 +
e^{2\Lambda(r)}dr^2+r^2d\theta^2 \nonumber\\
+ r^2\sin^2\theta\left(d\varphi-\omega dt\right)^2 \ ,
\label{eq:kerr}
\end{eqnarray}
where $e^{\Phi(r)}= e^{-\Lambda(r)}= \left(
1-{2GM/c^2r}\right)^{1/2}$ is the gravitational redshift function
of body (neutron star) of mass M, J is the angular momentum of a
neutron star and G is the gravitational constant. The metric in
equation (\ref{eq:kerr}) is the approximation of Kerr metric when
the angular momentum is small. The presence of the nondiagonal
component in metric in equation (\ref{eq:kerr})  results in the
well known effect of dragging of inertial frames of reference (the
Lense-Thirring effect)  with the angular velocity
\begin{equation}
\omega={2GJ\over c^2r^3} \ .
\end{equation}
Here $\tilde{\omega}=\omega -\Omega$ is the angular velocity of
the fluid as measured from the local free falling frame, $\Omega$
is the angular velocity of rotation of star relative to the
distant observer.

The stationary and locally nonrotating ``Zero Angular Momentum Observers''
(ZAMO)
(Bardeen et al 1972; Thorne \& Macdonald 1982; Thorne et al 1986)
are described by their four-velocity
\begin{eqnarray}
(u^{\nu})_{_{\rm ZAMO}}
\{\frac{1}{\sqrt{1-2M/r}},0,0,
\frac{\omega}{c\sqrt{1-2M/r}}\} \ , \nonumber\\
(u_{\nu})_{_{\rm ZAMO}}
\{-\sqrt{1-2M/r},0,0,0\} \ .
\label{eq:zamo}
\end{eqnarray}
The metric given by equations (\ref{eq:kerr}) has two Killing vectors
which are responsible for stationarity and axial symmetry and can be chosen
as
\begin{eqnarray}
\xi^\alpha_{(t)}=\left(1,0,0,0\right),\qquad
\xi^\alpha_{(\varphi )}=\left(0,0,0,1\right)\\
\omega=-\xi_{(t)}\cdot\xi_{(\varphi )}/\xi_{(\varphi)}\cdot
\xi_{(\varphi )} \ .
\end{eqnarray}
The four-velocity of ZAMOs is connected with the Killing vectors
according to formula
\begin{equation}
(u^{\alpha})_{_{\rm ZAMO}}
=\Lambda^{-1/2}\left(\xi^\alpha_{(t)}+\omega
\xi^\alpha_{(\varphi)}\right) \ .
\label{eq:kil}
\end{equation}
Write a steady, axisymmetric magnetic field of star as the sum of
a toroidal (i.e., azimuthal) component ${\mathbf B}^{\hat\varphi}$
and poloidal component ${\mathbf B}_p$ (which itself represents
the sum of the radial and axial components in cylindrical polars)
\begin{equation}
{\mathbf B}=B_{\hat\varphi}{\mathbf i}^{\hat\varphi} +{\mathbf
B}_p \ ,
\end{equation}
hats label the orthonormal components.

\begin{figure}
\includegraphics[width=0.35\textwidth]{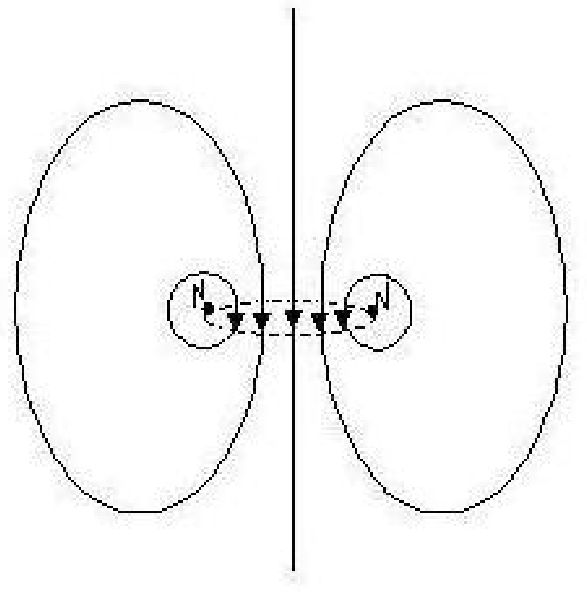}
\caption{\label{fig1} Magnetic force lines of the axial-symmetric
field in a meridional plane. Dashed line is the integration path
through the neutral points N.}
\end{figure}

Because of the axisymmetry, the magnetic configuration in all
meridional planes (through the axis of symmetry) is the same and
must consist of closed field lines (see figure~\ref{fig1}). In
each meridional plane there must therefore exist at least one
0-type neutral point (N), where a poloidal component of magnetic
field vanishes so that the field is purely azimuthal.
Now, the general relativistic Ohm's law in the form (\ref{eq:ohm's})
can be integrated around the closed line of force (C) through the neutral
points (N) to give
\begin{eqnarray}
\oint\frac{\Lambda^{1/2}}{\sigma}\jmath_\alpha dx^\alpha&=&
\oint\Lambda^{1/2}F_{\alpha\beta}u^\beta dx^\alpha\nonumber+ \oint
\Lambda^{1/2}R_{gg}\jmath^\beta A_{\alpha\beta} dx^\alpha\\&-&
\oint\Lambda^{1/2}R_H\left(F_{\alpha\sigma}+ F_{\rho\sigma}u^\rho
u_\alpha\right)\jmath^\sigma dx^\alpha \ . \label{eq:circ}
\end{eqnarray}
The component due to the gradient of pressure of charged particles
is curl-free and the circulation of the last term in the right
hand side of the equation (\ref{eq:ohm's}) vanishes\footnote {In
general, general relativistic Ohm's law contains contributions due
to the thermoelectric effects which are also curl-free.} if we
neglect the inhomogeneity of $N$. The terms on the right hand side
of equation (\ref{eq:ohm's}) describe, in order, the effects of
Ohmic decay, Hall effect and the effect of dragging of inertial
frames. If the second and third terms of the right-hand side of
(\ref{eq:ohm's}) can be neglected, this equation is reduced to the
standard form (having general-relativistic corrections due to
$\Lambda^{1/2}$) already used by many authors. Next we evaluate
the conditions under which the Hall current becomes important.
Comparing the rough magnitudes of the second term on the
right-hand side and the left hand side of (\ref{eq:circ}), we
cannot neglect the second term if magnetic field is
\begin{equation}
B\ge\frac{nec}{\sigma}\approx 1.44 \frac{n}{\sigma} G \ .
\end{equation}
However, the Hall drift does not work under a specific field
configuration, for example, if the toroidal field never appears,
the evolution of the pure poloidal field can not be influenced by
the Hall term.

One can evaluate the conditions when the gravitomagmetic term
becomes valuable. The rough comparison of the left-hand side and
the last term on the right-hand side of equation (\ref{eq:circ})
gives that the gravitomagnetic term cannot be neglected if
\begin{equation}
\omega\ge\frac{ne}{2m\sigma}\approx 2.7\cdot 10^{17}
\frac{n}{\sigma} s^{-1} \ .
\end{equation}
It shows that in principle the Hall drift is much stronger than
the gravitomagnetic one and in this respect more important for the
astrophysical processes. Only in the exceptional case when the
Hall drift does not work one can discuss the gravitomagnetic
effect.


The four-velocity of plasma can be decomposed in the form
\begin{eqnarray}
u^\alpha=\frac{u^\alpha_{(0)}+v^\alpha /c}{\sqrt{1-v^2/c^2}}
\approx u_{(0)}^\alpha+v^\alpha /c= \nonumber\\
(u^{\alpha})_{_{\rm ZAMO}}
-\Lambda^{-1/2}\omega\xi^\alpha_{(\varphi)}+ v^\alpha/c \ ,
\label{eq:decomp}
\end{eqnarray}
where four-velocity field
$u^\alpha_{(0)}=\Lambda^{-1/2}\xi^\alpha_{(t)}$ is parallel to the
timelike Killing vector.

Let us calculate the value of the integral
\begin{eqnarray}
\oint\Lambda^{1/2}F_{\alpha\beta}v^\beta dx^\alpha=
\oint\Lambda^{1/2}\left(E_\beta v^\beta\right)
(u_{\alpha})_{_{\rm ZAMO}}
dx^\alpha -
\nonumber\\
\oint\Lambda^{1/2}\left((u_{\beta})_{_{\rm ZAMO}}
v^\beta\right)
E_\alpha dx^\alpha
+\oint\Lambda^{1/2}e_{\alpha\beta\mu\nu}
(u^{\mu})_{_{\rm ZAMO}}
B^\nu v^\beta dx^\alpha
\label{eq:circ0}
\end{eqnarray}
when $\alpha=3$.

In this case the first integral is equal to zero according to
(\ref{eq:zamo}). From formulae (\ref{eq:kil}) and (\ref{eq:decomp}) one
can get
\begin{equation}
\Lambda^{1/2}\left((u^{\beta})_{_{\rm ZAMO}} v_\beta\right)=
\omega\xi^\beta_{(\varphi)}v_\beta\ .
\end{equation}
It is meant that in general due to `the dragging of the reference
frame' the second term in (\ref{eq:circ0}) does not vanish since
the electric field has two contributions: one is proportional to
the electric current and second one being produced by the stellar
magnetic field is proportional to angular velocity. Thus the last
one is also negligible in the linear in angular velocity of
rotation approximation.

The third integral
\begin{equation}
\oint\Lambda^{1/2}r\sin\theta \left(B^{\hat\theta} v^{\hat
r}-B^{\hat r}v^{\hat\theta}\right)d\varphi=0
\end{equation}
in the right hand side of (\ref{eq:circ0}) which is the induction
term due to the hydrodynamic motion of plasma with velocity
$v^\beta$ is identically vanishing since $B^{\hat
r}=B^{\hat\theta}=0$ along a contour through the neutral points
$N$.

By using of Maxwell equations (\ref{eq:max})
and Stoke's theorem (Misner et al 1973) one can show that
\begin{eqnarray}
\oint \Lambda ^{1/2} F_{\alpha\beta}u^\beta_{(0)} dx^ \alpha
=-\frac{1}{2}\int \left({\it\L}_{\xi_t}F_ {\alpha\beta}\right)
dS^{\alpha\beta} = 0 \ .
\end{eqnarray}
Since the electromagnetic field is stationary by assumption, i.e.
${\pounds}_{\xi_t} F_{\alpha\beta}=0$, then the first term under
the integral on the right-hand side of equation (\ref{eq:circ})
vanishes.

Thus the integral of Ohm's law~(\ref{eq:circ}) reduces to
\begin{eqnarray}
\oint\frac{\Lambda^{1/2}}{\sigma}\jmath_\alpha dx^\alpha=
-\oint\Lambda^{1/2}R_H\left(F_{\alpha\sigma}+ F_{\rho\sigma}u^\rho
u_\alpha\right)
\jmath^\sigma dx^\alpha+\nonumber\\
\oint\Lambda^{1/2} R_{gg}\hat\jmath^\beta A_{\alpha\beta} dx^\alpha +
\oint\Lambda^{1/2}\omega \left(\xi^\beta_{\varphi} v_\beta\right)
E_\alpha dx^\alpha \ .
\label{eq:cow}
\end{eqnarray}
Hence exact maintenance of the field in the principle is possible.
The last term on the right hand side of the equation
(\ref{eq:cow}) is the one first discussed by Khanna \& Camenzind
(1996). It disappears in the linear approximation in $\omega$ if
there are no any conduction currents in the plasma.

The physical interpretation of the other terms  on the right hand
side of the equation (\ref{eq:cow})  is as follows. On one hand
changes in the field are due to the motion, which transports the
field lines from point to point, and to the finite conductivity,
which permits the field to diffuse from point to point and so
decay. But the Hall and gravitomagnetic force effects on radial
current produce the current in azimuthal direction which creates
new field lines. Thus there is, in principle, the mechanism to
balance the diffusive decay of the field and a steady state is
possible. However the importance of the Hall and gravitomagnetic
effects depends on the model since these effects does not work
under a specific field configurations. For more details about the
effect of Hall drift in the neutron stars, see e.g. \cite{pg10},
\cite{pg07},  Jones (1988), \cite{rg02}, Goldreich \& Reisenegger
(1992), Naito \& Kojima (1994), Muslimov (1994).

This result means that the stationary axisymmetric electromagnetic field
in general relativity, in fact, can be supported by the relativistic
rate of rotation $A_{\alpha\beta}$. For example, for the plasma
embedded in the Schwarzschild space time with $A_{\alpha\beta}\equiv 0$
the classical Cowling's antidynamo theorem is valid. But for the
space-time of slow rotating compact object with nonvanishing nondiagonal
components of metric tensor, the relativistic rate of rotation
$A_{\alpha\beta}$ is nonzero. The radial current experiences effect of
the gravitomagnetic force and therefore according to (\ref{eq:ohm's})
we have the following value for the gravitomagnetically
generated azimuthal current
\begin{eqnarray}
\jmath^{\hat\varphi} =-\frac{R_{gg}\sigma \jmath^{\hat
r}\sqrt{1-2GM/c^2r}A_{\varphi r}}{r}=
\nonumber\\
\frac{R_{gg}\sigma\jmath^{\hat
r}}{c\left(1-2GM/c^2r\right)}\{aM/r^3\} \ , \label{eq:azimut}
\end{eqnarray}
where
\begin{equation}
A_{r\varphi}=\frac{aM}{cr^2\left(1-2GM/c^2r\right)^{3/2}} \ .
\end{equation}

The total current $I$ through $\varphi =const$ plane of plasma is
\begin{equation}
I=\int\jmath^{\hat\varphi}
r\left(1-2GM/c^2r\right)^{-1/2}drd\theta \ . \label{eq:total}
\end{equation}
One can use approximate behavior for azimuthal current during the
rough evaluations:
\begin{equation}
\jmath^{\hat\varphi}\approx\frac{R_{gg}\sigma j^{\hat r}\omega}{c}
\ .
\end{equation}
The main question to be answered is how strong the rotational
amplification of the electric current can become. For the typical
value of parameters $\Omega=10^3\textrm{s}^{-1}$, $N=10^7
\textrm{cm}^{-3}$, $\sigma=10^7 T^{3/2} \textrm{s}^{-1}$,
$m=9.1\times 10^{-28}\textrm{g}$, $e=4.8\times 10^{-10}
\textrm{cm}^{1/2}\cdot g^{1/2}\cdot \textrm{s}^{-1}$ and
$c=3\times 10^{10}\textrm{cm}\cdot \textrm{s}^{-1}$, the
dimensionless parameter ${R_{gg}\sigma\omega}/{c}$ can reach big
numbers if the temperature of plasma $T$ exceeds, for instance,
$10^{9}-10^{10} \textrm{K}$. Such temperatures are realized for
pulsar's plasma and in this connection the discussed mechanism of
generation of azimuthal current can produce the magnetic field
which will compensate the ohmic decay of magnetic field. The
temperature dependence is more stronger if the radial current is a
consequence of the temperature instabilities, that is if
$\jmath^{\hat r} =-\sigma\beta \textrm{grad} T$ since
thermoelectric power $\beta$ also depends on the temperature as
$T^{3/2}$.

Thus the result~(\ref{eq:cow}) leads to the deep understanding that even
very small radial currents in pulsar's plasma can be essentially amplified
by the gravitomagnetic force effects. Furthermore, from our
point of view, the observational evidence on the existence of
rotational effects on the conduction current provides the laboratory
experiment of Vasiliev (1994) where the vertical magnetic field around
rotating
cylindrical conductor with the radial current has been detected. The
experiment  has been theoretically explained (Ahmedov 1998) with help of the
general relativistic Ohm's law for conduction current which includes
rotational and gravitomagnetic terms.

Magnetic field created by the azimuthal current in a current loop
in the Schwarzschild and Kerr space times has been considered by
Petterson (1974) and Chitre \& Vishveshwara (1975), respectively.
For a current carrying loop located in the equatorial $\theta
=\pi/2$ plane, symmetrically around the slowly rotating star, at
radius b, the magnetic field in the region $r\ge b$ is given by
\begin{eqnarray}
B^{\hat r}
&=&-\frac{3\mu\cos\theta}{4M^3}\left[\ln\left(1-\frac{2M}{r}\right)+
\frac{2M}{r}\left(1+\frac{M}{r}\right)\right] \ ,\nonumber\\
B^{\hat\theta} &=&\frac{3\mu\sin\theta}{4M^2r}\left[\frac{r}{M}\ln
\left(1-\frac{2M}{r}\right)
+\left(1-\frac{2M}{r}\right)^{-1}+1\right]\nonumber\\
&&\times \left(1-\frac{2M}{r}\right)^{-1/2} \label{eq:field}
\end{eqnarray}
and identical to the magnetic field of a dipole with moment
\begin{eqnarray}
\mu=\pi b^2\left(1-2M/b\right)^{1/2}I \ .
\label{eq:moment}
\end{eqnarray}

As one can see from (\ref{eq:azimut}), (\ref{eq:total}),
(\ref{eq:moment}) the behavior of magnetic field (\ref{eq:field})
produced by an arbitrary azimuthal current strongly depends on the
amplification parameter.

\section{Conclusion}

We study the equations of motion of two-component plasma embedded
in the external gravitational field and derive a generalized Ohm's
law for a fully ionized electron-ion plasma within the framework
of general relativity. Then combining Maxwell's equations and
Ohm's law in a stationary and axisymmetric geometry we obtained
the approximate (asymptotic) equations that describe the external
electromagnetic field and electric current in a plasma (with
radial electric current) surrounding a rotating neutron star. By
considering a simplified model, we have found that gravitomagnetic
effect on radial current may produce increasing azimuthal current
in same special conditions in plasma, and therefore a dynamo in
principle may be possible. Our general conclusion from the above
analysis may be summarized as follows:

\noindent 1. The general relativistic Ohm's law for plasma
contains new terms as compared with special relativity case in the
limit of quasi-neutral plasma, which is caused by the
gravitomagnetic effects on the electric current.

\noindent 2. The azimuthal current arises from the gravitomagnetic force
effects on radial current in the plasma surrounding
rotating neutron star.

\noindent 3. The most important that due to this gravitomagnetic
effect the circulation of azimuthal current is not equal to zero
even in a steady state, which in principle  may allow axisymmetric
dynamo action in some special cases.

\noindent 4. Rough evaluations for magnetic field arising from the
induced azimuthal current for the typical values of parameters
give that its quantity may be comparable with the stellar magnetic
field only for exceptional cases when the plasma is rare and
simultaneously has high temperature. This result provides us a
right to say that Cowling's theorem, in general,
 can not be
violated in the metric of rotating neutron star by the new gravitomagnetic
terms in Ohm's law. Thus if the astrophysical object is a rotating
compact star, the gravitomagnetic potential and spin of the compact
star may, in fact, produce an axial symmetric magnetic field in the
surrounding plasma carrying conduction current but the mechanism of its
amplification for possible dynamo needs further more detailed investigation.

\section*{Acknowledgments}

I wish to thank Bobomurat Ahmedov and Viktoria Morozova for
bringing this problem to my attention and for many helpful
discussions and comments. I thank the IUCAA for the hospitality
where the research has been conducted.

\bibliographystyle{spr-mp-nameyear-cnd}

\end{document}